\documentclass[a4paper,10pt]{article}

\usepackage[a4paper,top=2cm,bottom=2cm,left=2.5cm,right=2.5cm]{geometry}

\usepackage[utf8]{inputenc}
\usepackage{fontenc}
\usepackage{graphicx}
\usepackage{url}
\usepackage{xspace}
\usepackage{color}
\usepackage[labelformat=simple]{subcaption}
\usepackage{indentfirst}
\usepackage{authblk}
\usepackage{units}
\usepackage{caption}
\DeclareCaptionLabelFormat{adja-page}{\hrulefill\\#1 #2 \emph{(previous page)}}
\usepackage{mathtools} 
\usepackage{amsmath}
\usepackage{amssymb}
\usepackage{commath} 
\usepackage{lineno} 
\graphicspath{{./images/}}

\title{Cross-inhibition leads to group consensus despite the presence of strongly opinionated minorities and asocial behaviour}
\author[1,2,*]{Andreagiovanni Reina}
\author[1]{Raina Zakir}
\author[3,4]{Giulia De Masi}
\author[3]{Eliseo Ferrante}
\affil[1]{Institute for Interdisciplinary Studies on Artificial Intelligence (IRIDIA), Universit\'{e} Libre de Bruxelles, Brussels, Belgium}
\affil[2]{Sheffield Robotics, University of Sheffield, Sheffield, UK}
\affil[3]{Technology Innovation Institute, Abu Dhabi, UAE}
\affil[4]{BioRobotics Institute, Sant'Anna School of Advanced Studies, Italy}
\affil[*]{andreagiovanni.reina@gmail.com}
\date{}

\begin{document}
\maketitle

\section*{Abstract}
  Strongly opinionated minorities can have a dramatic impact on the opinion dynamics of a large population. Two factions of inflexible minorities, polarised into two competing opinions, could lead the entire population to persistent indecision. Equivalently, populations can remain undecided when individuals sporadically change their opinion based on individual information rather than social information. Our analysis compares the cross-inhibition model with the voter model for decisions between equally good alternatives, and with the weighted voter model for decisions among alternatives characterised by different qualities. Here we show that cross-inhibition, contrary to the other two models, is a simple mechanism
  that allows the population to reach a stable majority for one alternative even in the presence of a relatively high amount of asocial behaviour.
  The results predicted by the mean-field models are confirmed by experiments with swarms of 100 locally interacting robots. This work suggests an answer to the longstanding question of why inhibitory signals are widespread in natural systems of collective decision making, and, at the same time, it proposes an efficient mechanism for designing resilient swarms of minimalistic robots.

\section*{Introduction}

How a group of individuals reaches a consensus through local interactions and without any centralised authority is widely studied in social, biological, political, and information sciences \cite{Castellano2009, Conradt2009, Baronchelli2018, seeley2011, Valentini2017, Reina:SwInt:2021}.
The voter model \cite{Clifford1973, Holley1975}, thanks to its simplicity and tractability, has been
employed both to describe a large variety of natural systems (from plants to humans) across all scales of biological complexity \cite{Jhawar2020, FernandezGracia2014, Zillio2005, Redner2019}, and to engineer decentralised artificial systems, such as robot swarms \cite{Valentini2017}. Recent studies have however shown that the consensus dynamics of the voter model can easily be jeopardised, by even a minimal number of self-willed individuals \cite{Mobilia2007, Khalil2018, Galam2007}.
%
For each alternative, the presence of just one inflexible individual (also called zealot or stubborn in literature), influencing others but not changing its opinion, prevents the entire population from reaching any stable agreement (i.e., a large majority with the same opinion) \cite{Mobilia2007}.
Equivalently, in the absence of zealots, the population cannot achieve a stable majority if every individual makes spontaneous (asocial) changes of opinion, even if sporadic (modelled as the noisy voter model \cite{Khalil2018}).

\begin{figure}[t!]
  \includegraphics[width=\textwidth]{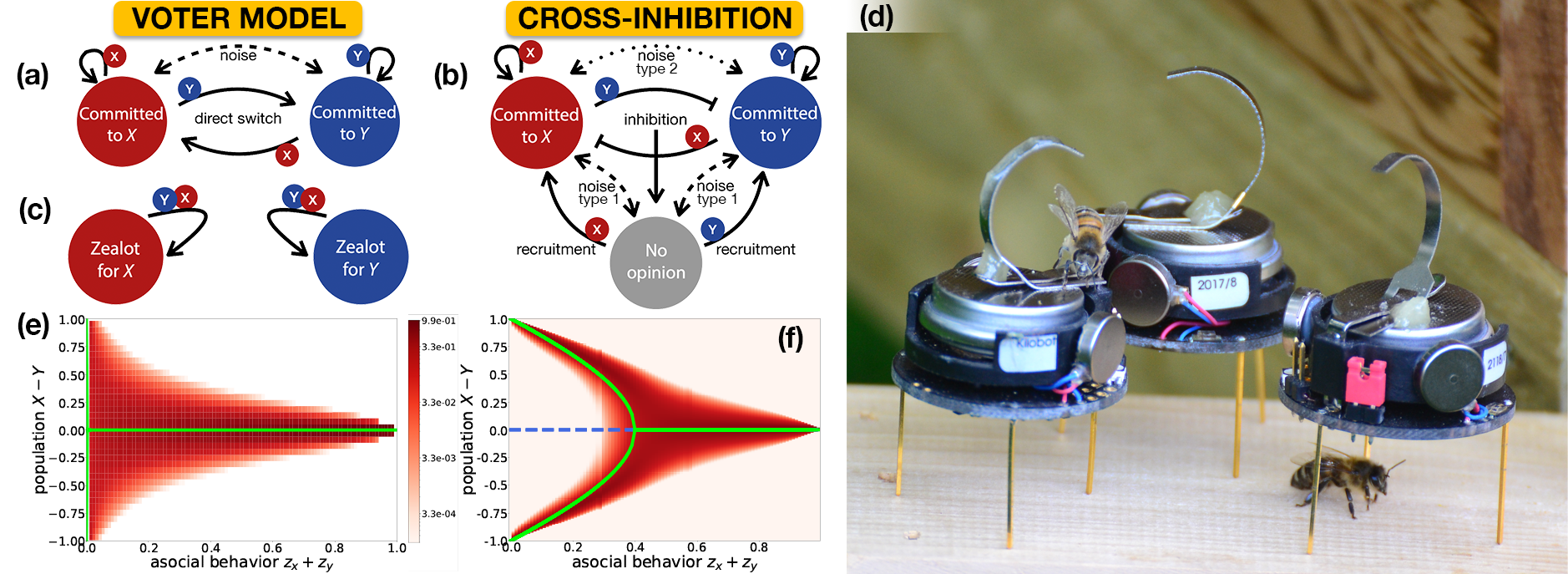}
  \caption{\textbf{In nature and robotics, populations of simple individuals can make consensus decisions through the cross-inhibition model.} \textbf{(a-b)}~Motifs of the voter model and the cross-inhibition model. The commitment of an individual (in circles) changes upon interaction with peers (illustrated by solid arrows with a small circle that indicates the peer's commitment). In the voter model, individuals directly switch their commitment between the two alternative opinions X and Y. Instead, in~(b), individuals with different opinions cross-inhibit each other, and the inhibited individual reverts to an uncommitted state. Individuals without any opinion can be recruited by committed peers. Both models can be subject to noise (dashed arrow), that corresponds to an asocial switch of opinion (independent of peers) which can be caused by self-sourcing information. Noise in the cross-inhibition model can be of two alternative types (dashed or dotted arrows).  \textbf{(c)}~Zealots are inflexible and never change their opinion upon receiving neighbours' votes. \textbf{(d)}~Biological inspiration can be employed to design safe robotic systems able to reach a stable consensus. The cross-inhibition model has been first introduced to describe house-hunting in honeybees~\cite{Seeley2012}. We used the biological model to design a resilient collective behaviour in a swarm of 100 Kilobot robots. Kilobots are simple robots widely used for collective intelligence studies~\cite{Rubenstein2014}. \textbf{(e-f)}~Bifurcation analysis of the mean-field ODE models can be used to identify when a large population can reach an agreement or remains split into two polarised factions. We consider the presence of two groups of zealots, equally split between the two opinions. We display the fixed points of Eqs.~(\ref{eq:ode:vm:zealots:full}) and (\ref{eq:ode:ci:generic}) (solid green lines are stable and dashed blue are unstable) in overlay to the stationary probability distribution of the master equation for a population of $S=200$ individuals (red colour-maps computed using the master equation's analytical solution of Eqs.~\eqref{eq:spd:vm:diff}, \eqref{eq:spd:ci:diff}). (e)~Using the voter model, the population gets locked into an indecision state as soon as a minimal number of zealots (x-axis) are introduced. (f)~Instead, using the cross-inhibition model, the population demonstrates resilience against relatively high levels of zealotry. For values of asocial behaviour smaller than the supercritical pitchfork bifurcation, the system has two stable fixed points representing population agreement for either option. Instead, after bifurcation (when zealots comprise more than 40\% of the entire population), the system with high levels of asocial behaviour has a single attractor, representing indecision: a deadlocked population unable to make any collective decision.}
  \label{fig:motifs}
\end{figure}

Here, we show that a model of comparable simplicity, the cross-inhibition model \cite{Seeley2012, Reina:PLOSONE:2015, Reina:PRE:2017}, makes the population resilient to the presence of both self-willed and inflexible individuals, and able to reach a stable consensus. These results contribute to explaining why inhibitory signals evolved in biological systems that need group consensus despite operating in noisy contexts, such as social insect colonies and neuronal populations \cite{Seeley2012,Bogacz2006}. Additionally, our findings are relevant to the design of minimal interaction patterns that allow social networks and decentralised robotic systems to reach an agreement, despite the presence 
of noise or the asocial behaviour of some individuals, whether unintentional or malicious \cite{Higgins2009,DeMasi2021}.
In both the voter and cross-inhibition models, population consensus emerges from pairwise interactions between individuals, with a randomly selected speaker and listener.
The main difference between the two models is that when the speaker and the listener have different opinions, in the voter model the listener directly switches to the speaker's opinion, while in the cross-inhibition model the listener becomes undecided or, in other words, gets inhibited and remains without an opinion. Only undecided individuals switch to the speaker's opinion (see Figure \ref{fig:motifs}a-b).

We model asocial behaviour as noise or zealotry, which have both been shown to be mathematically equivalent in the macroscopic-level model \cite{Khalil2018}. Instead, at the microscopic level, the individual asocial behaviour is distinct in case of noise and zealotry. Through noise, every individual sporadically changes its opinion independently of others' opinions. Zealots always ignore others' opinions and remain unmovable (Figure~\ref{fig:motifs}c); a form of antisocial behaviour more extreme than noise that however is followed by only a minority of the population \cite{Galam2011,Galam2020}.
Individuals may have asocial behaviour 
because they fail to comply with social pressure, or retrieve information from independent sources other than peers \cite{Zakir:ANTS:2022}. Either situation is common in several biological and artificial systems that rely on noisy communication and operate in complex environments with abundant sources of conflicting information. Identifying mechanisms that provide resiliency and stability to the group is key to understanding how living groups can achieve social homeostasis \cite{holldobler1990ants}. The cornerstone mechanism of the cross-inhibition model is inhibitory signalling between individuals with opposing opinions. This mechanism is widespread in nature. Evolution has recurrently reached the use of inhibition as the most efficient method to `design' interactions of complex systems. Inhibition can be found in apparently very different systems, such as in cell metabolism \cite{Cardelli2017}, neuronal activity \cite{Bogacz2006}, honeybee house-hunting \cite{Seeley2012} and human societies \cite{Franci:SwInt:2021}, however the underlying models can have striking similarities in their structure and dynamics \cite{Cardelli2017, Marshall2009, Reina:scirep:2018, Borofsky2020}. Previous studies have shown that cross-inhibition can break the symmetry in absence of asocial behaviour \cite{Seeley2012,Reina:PRE:2017}. Our analysis reveals an additional key feature of the minimal cross-inhibition model: the ability to reach a stable consensus despite asocial behaviour.
This feature is pivotal in biological systems seeking to achieve coordinated actions, for example, house-hunting honeybees \cite{Seeley2012}, neurons firing to discriminate between stimuli \cite{Bogacz2006}, and individuals agreeing on social norms \cite{Baronchelli2018}. Identifying minimal mechanisms capable to lead to a stable consensus despite asocial behaviour is also relevant in shielding artificial systems from deficient behaviour caused by malfunction or cyber-attacks (Figure~\ref{fig:motifs}d).

%

\section*{Results and Discussion}


Through mean-field analysis, we show that the ODE system describing the cross-inhibition model predicts a stable group consensus (two stable fixed points)
despite a considerable proportion of zealots, or, equivalently, a high level of noise. More precisely,
when zealots comprise less than 40\% of the entire population or the noise is below a given threshold  (i.e. prior to the supercritical pitchfork bifurcations in Figures~\ref{fig:motifs}f, \ref{fig:noise}e, and \ref{fig:noise}g), the ODE system has two stable fixed points representing population agreement for either option.
Instead, the voter model falls in a permanent undecided state as soon as minimal asocial behaviour is present (single green attractor representing indecision in Figures~\ref{fig:motifs}e, \ref{fig:noise}a, and \ref{fig:noise}c).
%
However, previous research has shown that, in small populations, the voter model can break the symmetry (i.e. the decision deadlock) induced by fluctuations of a finite-sized system \cite{Biancalani2014, Herrerias-Azcue2019}. Additionally, the stable points of the cross-inhibition model are not absorbing states---that is, there is always the possibility that through random fluctuations the system changes state and moves from the basin of attraction of a stable point to the other.  
Therefore we conduct semi-analytical and computational analyses of the master equation of the two models in order both to study the stochastic dynamics of finite-sized systems and to quantify the stability of the fixed points.

\subsection*{Finite-sized system analysis}

\begin{figure}[t!]
  \centering
  \includegraphics[width=\textwidth]{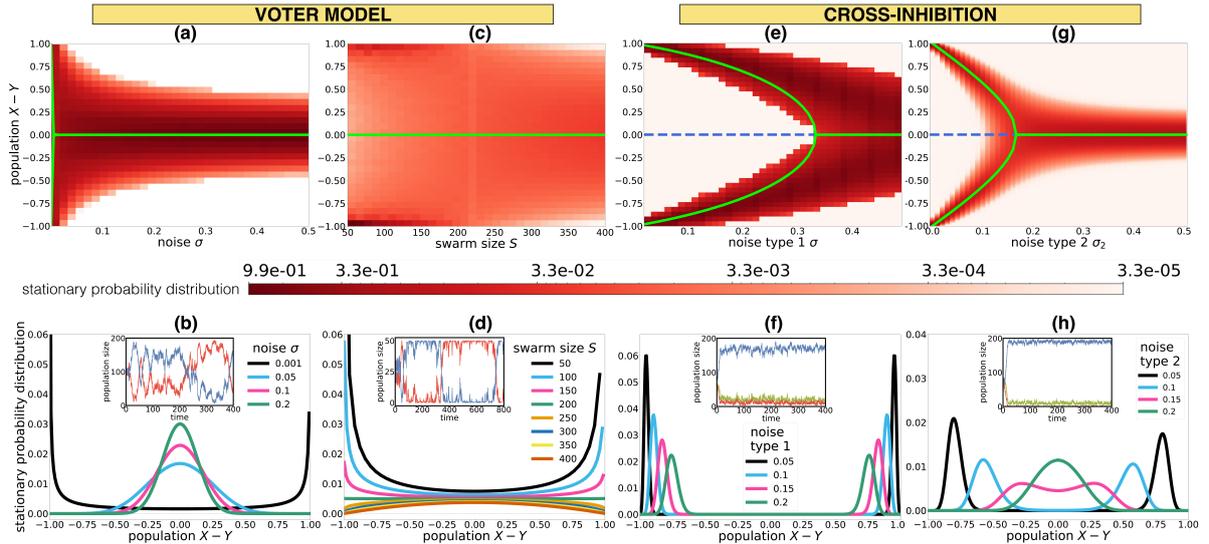}
  \caption{\textbf{The voter model cannot break the symmetry for increasing levels of asocial behaviour or swarm size, instead the cross-inhibition model can.} On the top row, the red colour-maps show the stationary probability distribution (SPD)---computed via the master equations, except (g)-(h) which are computed with simulations---of a swarm with $S=200$ individuals to be in the collective state indicated on the y-axis (see Methods). The overlying lines show the fixed points of the ODE system (solid green are stable, dashed blue are unstable). On the bottom row, each of the main plots shows a number of SPDs that are in correspondence with the colour-map on the top. The insets are a trajectory of the change of commitment over time for a representative simulation. Panels~(a-d) are computed through Eq.~\eqref{eq:spd:vm:diff}, panels (e-f)~through Eq.~\eqref{eq:spd:ci:diff}, and (g-h)~computationally using $10^4$ simulations of Gillespie's stochastic simulation algorithm.  \textbf{(a-d)}~As previously reported \cite{Biancalani2014, Herrerias-Azcue2019, Khaluf2017}, for small levels of noise and small swarms, despite different predictions of the mean-field ODE system, the voter model is often in a state of collective agreement. However the majority is slim and the dynamics highly unstable (e.g., see insets of (b) and (d) for $\sigma=0.005$).  Additionally, the population rapidly goes to indecision as soon as the noise (in a-b, with $S=200$) or the system size (in c-d, with $\sigma=0.005$) increases: the two symmetry-breaking peaks, for -1 and +1, transition into a single peak at $x-y=0$, that represents indecision. \textbf{(e-h)}~Differently, the cross-inhibition model is able to break the symmetry and reach an agreement despite relatively high levels of noise---that is, asocial behaviour of individuals that spontaneously change opinion. We report the results for both types of noise (see x-axis and Figs.~\ref{fig:motifs}e-g) and for different swarm sizes (in Supplementary Figure~S1). Increasing noise shifts the agreement to values lower than full consensus. Nevertheless, prior to bifurcation, a large majority for one opinion is maintained with very stable dynamics (e.g., see insets of f and h for $\sigma=0.05$ and $S=200$, tenfold the noise used in insets of b and d).}
  \label{fig:noise}
\end{figure}

While the mean-field ODE system of the voter model predicts a single stable point with a polarised population split between the two options, Figures~\ref{fig:noise}a-d show that a small-sized system with moderate levels of noise is most often in a state of agreement for either option.
However, noise-induced bistability leads to highly unstable consensus states (e.g. see the insets of Figures~\ref{fig:noise}b,d), and, in addition, is vulnerable to an increase in both the system size $S$ and the noise level $\sigma$. As shown by the master equation analysis in Figures~\ref{fig:noise}a-d, for noise values or population sizes greater than a threshold $\sigma^{-1}=S$, the voter model transits from a regime of decision to a regime of indecision \cite{Biancalani2014, Herrerias-Azcue2019}. The impact of zealotry on consensus is even more accentuated. Two equally-sized and relatively small groups of inflexible zealots are sufficient to lock a much larger social population into a state of indecision. While in the voter model a small level of asocial behaviour impedes population agreement, on the contrary the cross-inhibition model breaks the symmetry sustained by large proportions of zealots or by high levels of noise 
(Figures~\ref{fig:motifs}f and \ref{fig:noise}e-h). 
For the cross-inhibition model, we consider two alternative types of noise (dashed and dotted lines in Figure~\ref{fig:motifs}b). Through noise type~1, individuals do not immediately change opinions but go through the undecided state before adopting a new opinion (following the inhibitory mechanism). The behaviour of noise type 1 is analogous to the way in which individuals react to social information received from other individuals, including zealots.
Instead, noise type~2 corresponds to the way noise is implemented in the voter model with individuals directly switching their opinions (e.g., due to self-sourced environmental information) and allows a more thorough comparison of the two models. For both types of noise (Figure~\ref{fig:noise}e-h) and for any tested swarm size (see Supplementary Figure~S1), the cross-inhibition model has qualitatively the same dynamics: it breaks the symmetry despite relatively high levels of social noise.

%

\subsection*{Swarm robotics experiments}

\begin{figure}[t]
  \centering
  \includegraphics[width=\textwidth]{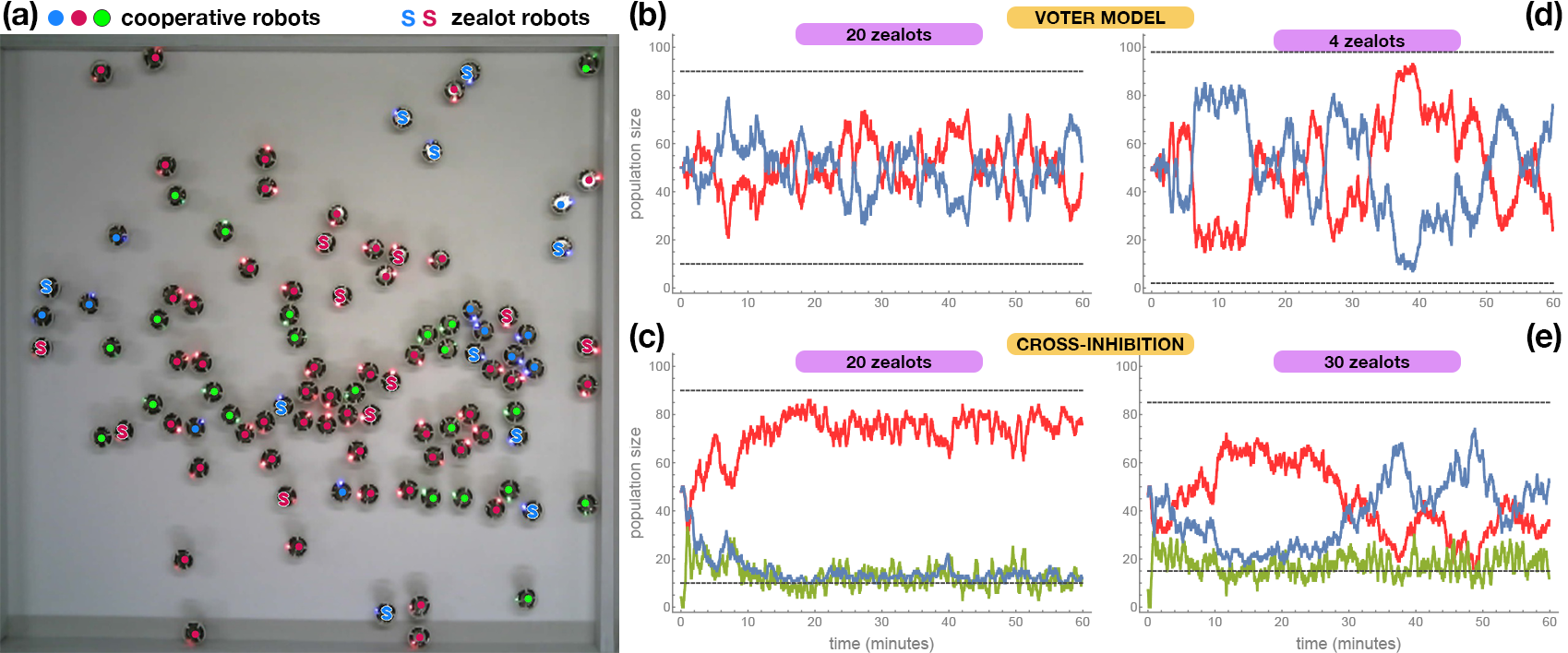}
  \caption{\textbf{We implemented two models on a swarm of 100 Kilobot robots that display their opinion---red or blue---via their colour LED; green robots have no opinion.} \textbf{(a)} The Kilobots~\cite{Rubenstein2014} move through vibration motors in a flat environment and exchange infrared messages every 30 seconds with one another in a range of about \unit[10]{cm}. Upon receiving a message they change their opinion according to one of the two tested models: the voter and the cross-inhibition model. We included a number of zealot robots that only broadcast but never read other robots' messages, thus never change opinion. We split the zealots equally between the red and blue options. The displayed image was taken with an overhead camera midway through an experiment. Videos are available as Supplementary Movies~2-5. In panels \textbf{(b-e)}, we report the change over time of the number of robots with different commitment states, the red and blue curves are bounded by the two horizontal dashed lines determined by the number of inflexible zealot robots. (b)~Using the voter model, the swarm is unable to reach a decision when 20 of the robots are zealots. (c)~Instead, the cross-inhibition model is able to quickly converge to a consensus and maintain it for the entire length of the experiment. (d)~Just 4 zealots---that is, two inflexible robots for each option---are enough to destabilise the agreement of a swarm of 96 robots that cooperate through the voter model. (e)~The cross-inhibition model can also suffer instability when a large proportion of robots are zealots. In this experiment, there are 30 zealots and 70 cooperative robots.}
  \label{fig:robots}
\end{figure}

We tested our theoretical results through a set of experiments with 100 Kilobot robots that locally interact with each other in order to reach an agreement (see Figure~\ref{fig:robots}a and Supplementary Movie~1, also available at \url{https://youtu.be/mQtLhMqdVWg}). The swarm starts from a polarised indecision, with a 50-50 split for either alternative (represented by the blue and red colours).
To test the resiliency of the two models against asocial behaviour in the robot swarm, we included 20 zealot robots---equally split between the two options---that only broadcast their opinion but do not listen to others. Figure~\ref{fig:robots}b shows that the remaining 80 robots updating their commitment using the voter model are unable to reach any agreement within one hour. Instead, through the cross-inhibition model, the swarm rapidly selects any of the two alternatives and maintains a stable agreement (Figure~\ref{fig:robots}c). We can further appreciate the fragility of the voter model in the experiment of Figure~\ref{fig:robots}d where only four zealots---that is, two robots for each option---have been included in a swarm of 96 cooperative robots. The swarm remains undecided for the most part of the experiment and, in contrast to the cross-inhibition experiment with 20 zealots, the agreement is highly unstable. In Figure~\ref{fig:robots}e, we show that, in agreement with theory predictions, the stability of the cross-inhibition model can also be undermined by excessively large factions of zealots, in this case 30 zealots. The swarm of 70 cooperative robots is able to frequently attain a large majority for either alternative, however the decision is not stable and, over one hour, the consensus vacillates more than once.
Our swarm robotics experiments strengthen the validity of the theoretical models by showing that our mathematical equations can predict the collective behaviour of a swarm of 100 autonomous robots that locally interact with each other. Obtaining such correspondence was not obvious and corroborates the possibility of employing our theoretical results to implement resilient swarms of minimalistic robots.

\subsection*{Accuracy and reward in the best-of-n problem}

\begin{figure}[t!]
  \centering
  \includegraphics[width=\textwidth]{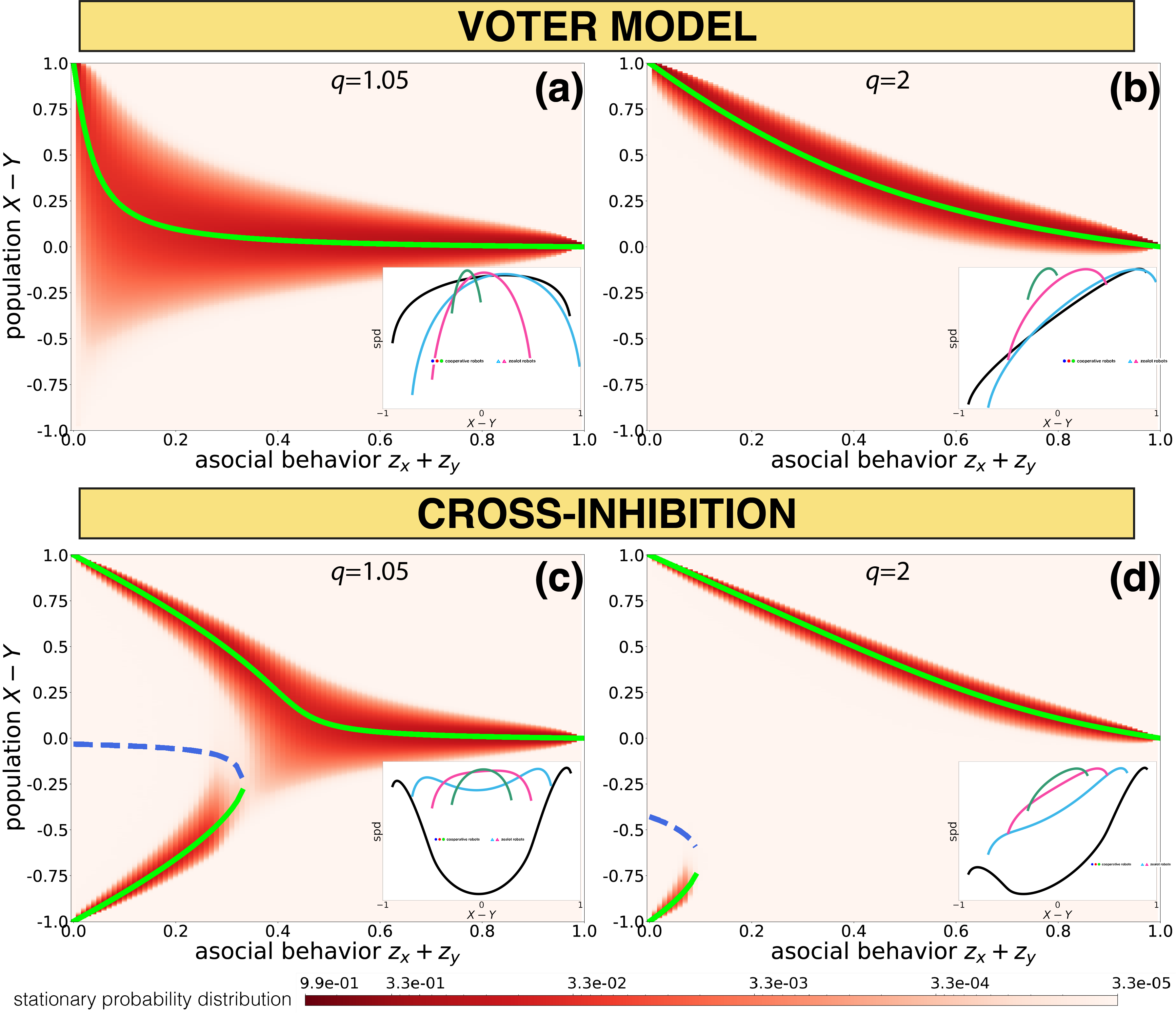}
  \caption{\textbf{Long term dynamics for decisions between options with asymmetric qualities (quality ratio $q=q_x/q_y\neq 1$) by populations with an infinite size (ODEs) and a finite size of $S=200$ individuals (master equation).} \textbf{(a-d)}~The main plot shows the ODEs' fixed points (solid green lines are stable and dashed blue lines are unstable) in overlay to the Gillespie's stochastic simulation algorithm (red heat-map, $10^4$ runs per condition) for increasing level of symmetric asocial behaviour (i.e., number of zealots equally split among the two options $z=z_x=z_y$). The inset of each panel shows the stationary probability distribution in log-scale computed with the analytical solution of the master equation at equilibrium, as from Eqs.~\eqref{eq:spd:vm:diff} and~\eqref{eq:spd:ci:diff}. \textbf{(a)}~The weighted voter model fails to reach a stable consensus when the two options have a similar quality, $q=1.05$: the population is either highly unstable (large fluctuations with few zealots) or undecided (polarised population with numerous zealots). \textbf{(b)}~Only for a higher quality ratio, $q=2$, the weighted voter model has a stable and large majority. \textbf{(c-d)}~Through cross-inhibition, the population can make a decision, despite the presence of a relatively large proportion of zealots, in any condition, both for small (c, $q=1.05$) or large (d, $q=2$) quality differences. However, for small quality differences (c), the high stability of the cross-inhibition model prevents the population to switch decision and can maintain a large consensus for the option with inferior quality. A stable majority for the inferior quality vanishes when either the quality difference is large (e.g., in d for $q=2$) or, counterintuitively, when the proportion of zealot increases.}
  \label{fig:qratio}
\end{figure}

\begin{figure}[t!]
  \centering
  \includegraphics[width=\textwidth]{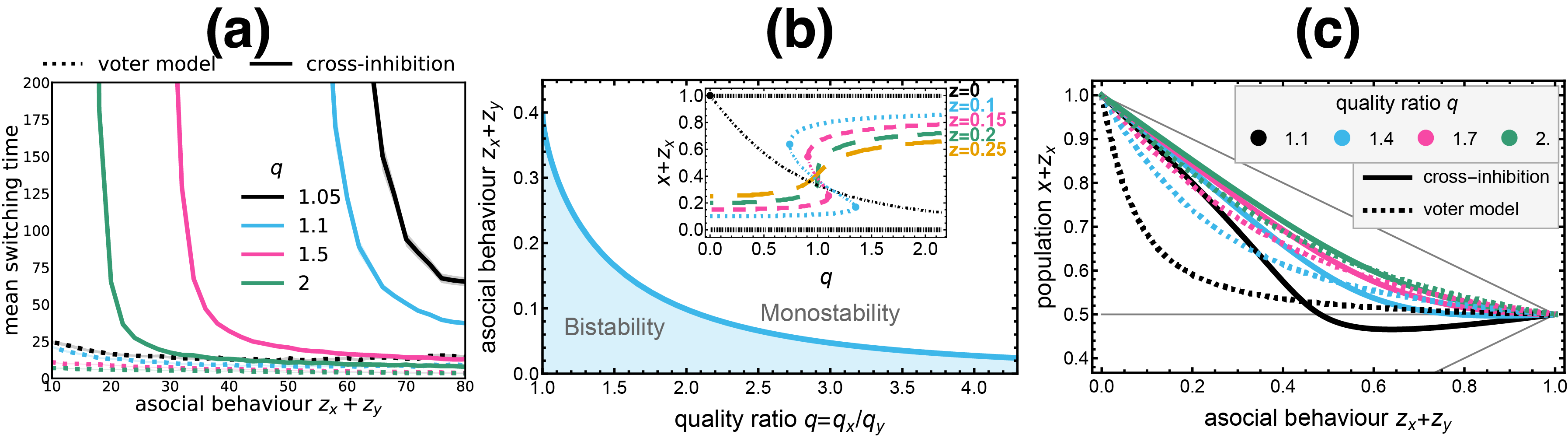}
  \caption{\textbf{Mean switching time and stability analysis for collective decisions between options with asymmetric qualities (quality ratio $q=q_x/q_y\neq 1$).} \textbf{(a)}~The mean switching time (MST) indicates the mean time necessary to reach a stable majority for the best option, when the population is initialised with a consensus for the option with lower quality (500 runs of the SSA, 95\% confidence interval indicated with colour-shades that are often smaller than the line-width). The voter model can quickly overturn the consensus for the inferior option. Instead, using the cross-inhibition model, the population is more rarely able to change its decision in favour of the best option in a reasonable time. The MST has low values when a large number of zealots are present, or when the quality difference is high. In the other cases, the cross-inhibition locks a large majority into a stable consensus for either option, which can also be the inferior one. \textbf{(b)}~The fold bifurcation point of the cross-inhibition model shows that by increasing asocial behaviour, the system transits from a phase of bi-stability to a phase with a single stable point. This phase transition threshold decreases with increased quality ratio $q$. The inset shows that increasing $z$ removes bistability but also makes the majority smaller. \textbf{(c)}~Nevertheless, cross-inhibition has always a larger majority than the weighted voter model, except for extremely high proportions of zealots, here illustrated as the stable fixed point of populations $x+z_x$ from the ODEs systems of Eqs.~(\ref{eq:ode:vm:zealots:full}) and~(\ref{eq:ode:ci:generic}).}
  \label{fig:bifurcation}
\end{figure}

In robotics, inspired by social insects' behaviour \cite{Marshall2009, Seeley2012}, decentralised decision models have been investigated in the context of the \textit{best-of-n} problem \cite{Valentini2016, Reina:DARS:2016, Talamali:SciRobot:2021}. The opinion of the individuals refers to an option which has a certain quality. While individuals make estimates of the quality that are subject to noise and often inaccurate, the group is able to agree on the option that is better evaluated on average by the population. Therefore, in the best-of-n problem, the goal is not only to reach a consensus but also to agree on the best option. To solve the best-of-n problem, the voter model has been generalised into the weighted voter model \cite{Valentini2014}. Individuals vote at a rate proportional to the estimated quality. The individuals directly switch their opinion in response to others' votes, following the same finite-state machine of the voter model of Figure~\ref{fig:motifs}a. As votes are more frequent for better options, the result is that the population reaches a consensus on the option that individuals have self-estimated to be the best.
By relying on the same strategy of quality-proportional voting, the cross-inhibition model has also been presented in a generalised form to tackle the best-of-n problem in artificial distributed systems \cite{Reina:PLOSONE:2015}.
Comparing the two models from the point of view of the best-of-n problem, we find that the two models maximise different metrics (Figure~\ref{fig:qratio}). The weighted voter model demonstrates higher accuracy, that is, the ability to select the best option, while cross-inhibition can yield a higher reward (as discussed after). Figure~\ref{fig:bifurcation}a illustrates the ability of the weighted voter model to always converge to the best option in a relatively short time, despite the system being initialised with a consensus for the inferior option. In real systems, random fluctuation may lead the population to an agreement in favour of the lower-quality alternative. Through the weighted voter model, inaccurate decisions can be reverted. However, especially when the options' qualities are similar,
the dynamics of the voter model are highly unstable (e.g. see the high spread of the red shade in Figure~\ref{fig:qratio}a) and can only grant a marginal majority for the best option (see also Figure \ref{fig:bifurcation}c).
Instead, the cross-inhibition model, by granting more stability to the collective agreement, can lessen accuracy by locking the population in a consensus in favour of the lower-quality alternative (Figure~\ref{fig:bifurcation}b).

Finding that the use of inhibition reduces collective accuracy is in contrast with extensive literature that shows inhibition as a pervasive mechanism in most natural systems that seek collective consensus~\cite{Cardelli2017, Bogacz2006, Seeley2012, Bizyaeva2020, Marshall2009}. Such discord can be reconciled by looking at a different metric. Rather than maximising accuracy, the cross-inhibition model looks better suited to yielding a high reward rate \cite{Pirrone2014}. When measuring accuracy, only the selection of the highest quality option is labelled as the correct (accurate) decision, while selecting any other option is incorrect. Instead, in value-based decisions,
the magnitude of the reward received by the decision-makers is used to measure the value of the chosen alternative.
On the one hand, considering accuracy can simplify the results' analysis \cite{Valentini2016, Talamali:ICRA:2019}. On the other hand, reasoning in terms of value-based decisions---in which the group benefits from the reward of the chosen option regardless of whether it is the absolute best in the environment---is more appropriate in most applications, both in engineered and biological systems (e.g., nutritional value during foraging, site quality during house-hunting, or aggregation location for the robot swarm) \cite{Pirrone2014, Bose:COBS:2017, Pirrone:TiCS:2021, Rajendran2022}.
As recent work has shown, the optimal strategy for value-based decisions can differ from normative accuracy-based decision policies \cite{tajima2016optimal}, especially in decisions between options with equal qualities \cite{Pirrone:TiCS:2021}.
Our analysis shows that, in line with the optimal policy \cite{tajima2016optimal}, the cross-inhibition model converges to the highest quality option when the quality difference between options is large (Figure~\ref{fig:qratio}d), and it trades accuracy for stability when the options have similar qualities (Figures~\ref{fig:qratio}c and~\ref{fig:bifurcation}a).
%
%
%

\subsection*{The benefits of inhibitory signals}

This study shows the inability of the voter model---both classical and weighted---to reach group coordination when symmetric noise or asocial behaviour is present, even in minimal quantities. In both artificial and natural systems, it is well justified to assume individuals may not always follow social pressure and sometimes act independently due to, for example, information acquired through individual exploration \cite{Zakir:ANTS:2022} or sensorial failures \cite{Tsimring2014}. The success of the voter model is due to its mathematical simplicity and tractability, however, for scientific progress, the fields of opinion dynamics and sociophysics require their models to include more aspects from real systems \cite{Schweitzer2018}. The cross-inhibition model is an alternative to the voter model that includes a widespread mechanism in nature: inhibitory signalling between individuals with different opinions. Such a simple change in individual behaviour revealed a dramatic change in the collective dynamics of the swarm, enabling system stability and symmetry-breaking.
Mathematically, the cross-inhibition model has similarities with the binary naming game \cite{Baronchelli2018} which is also capable of breaking decision deadlocks in the presence of symmetric inflexible populations \cite{Verma2014}. However, the rules of the naming game can be extended to problems with a larger number of options in a way that is fundamentally different from the cross-inhibition model, and this makes its analysis difficult due to the combinatorial explosion of the number of states \cite{Waagen2015}. In this article, we study the minimal formulation of a model that includes cross-inhibition and benefits from mathematical simplicity, as also shown in previous analytical studies that investigated the scalability of the cross-inhibition model to higher dimension problems \cite{Reina:PRE:2017}.
%
Our analysis shows that, with cross-inhibition, the collective performance is enhanced, rather than hampered, by noise (or zealots). Figure~\ref{fig:qratio} shows that increasing asocial behaviour to moderate values removes the possibility of inaccurate decisions while still granting a consensus decision. This finding is in agreement with previous research that showed that the emergence of group coordination, in humans, animals, or robots, is facilitated by noise \cite{Jhawar2020, Shirado2017, Rausch:SwInt:2019}. This study goes in the direction of building opinion dynamics models that gradually include more biologically plausible mechanisms while preserving mathematical simplicity.

\section*{Methods}

\subsection*{Chemical reactions}

Both models, illustrated in the motifs of Figures~\ref{fig:motifs}a-b can be written as a set of chemical reactions. The chemical reaction defines transition rates between the states in which an individual can be in. In our model, we define the following five states: commitment for option $X$ and $Y$, no opinion $U$, and zealots in favour of either option, $Z_X$ and $Z_Y$.

The voter model reads as:
\begin{equation}
  \begin{array}{lll}
    X + Y \xrightarrow{ q_x } X + X
    &\qquad \qquad  X + Y \xrightarrow{ q_y } Y + Y
    &\qquad \qquad \text{(direct switching)}\\
    Y \xrightarrow{ \sigma } X
    &\qquad \qquad   X \xrightarrow{ \sigma } Y
    &\qquad \qquad \text{(noise)} \\
    Y + Z_X \xrightarrow{ q_x } X + Z_X
    &\qquad \qquad X + Z_Y \xrightarrow{ q_y } Y + Z_Y
    &\qquad \qquad \text{(zealots)}.\\
  \end{array}
  \label{eq:chemical:vm}
\end{equation}
The rates of transitions resultant from interactions between individuals with different commitment (labelled in Eq.~(\ref{eq:chemical:vm}) as direct switching and zealots), are proportional to the option's qualities $q_x$ and $q_y$, for recruitment to option $X$ and $Y$, respectively. When noise is absent $\sigma=0$, there are no zealots $Z_X=Z_Y=0$, and the qualities are equal $q=q_x/q_y=1$, the model reduces to the classical voter model \cite{Clifford1973, Holley1975}. The system with $q \neq 1$ is a generalisation of the voter model in which individuals modulate their voting probability as a function of the option's quality, it has been named the ``weighted voter model'' \cite{Valentini2014} and has been introduced to make collective decisions that solve the best-of-n problem. Including noise $\sigma > 0$ into the voter model has been studied as the ``noisy voter model'' \cite{Khalil2018}. Noise can represent any form of independent (asocial) behaviour that leads the
individual to change its opinion, for example, self-sourced information. Therefore, the results from our model analysis can aslo be relevant to study how the frequency of using personal and social information, $\sigma$ and $(q_x+q_y)$, impacts the system dynamics. Zealots never change their opinion but influence the opinion of others \cite{Mobilia2003, Mobilia2007}. The impact of these individuals on the voter model has been shown to be equivalent to the impact of independent behaviour (i.e., noise), see \cite{Khalil2018} and our discussion below.

The cross-inhibition model reads as:
\begin{equation}
  \begin{array}{lll}
    X + Y \xrightarrow{ q_x } X + U
    &\qquad \qquad  X + Y \xrightarrow{ q_y } U + Y
    &\qquad \qquad \text{(cross-inhibition)}\\
    U + X \xrightarrow{ q_x } X + X
    &\qquad \qquad  U + Y \xrightarrow{ q_y } Y + Y
    &\qquad \qquad \text{(recruitment)}\\
    Y \xrightarrow{ \sigma } U \qquad U \xrightarrow{ \sigma } Y
    & \quad X \xrightarrow{ \sigma } U \qquad  U \xrightarrow{ \sigma } X
    &\qquad \qquad \text{(noise type 1)} \\
    Y \xrightarrow{ \sigma } X
    &\qquad \qquad   X \xrightarrow{ \sigma } Y
    &\qquad \qquad \text{(noise type 2)} \\
    Y + Z_X \xrightarrow{ q_x } U + Z_X
    &\qquad \qquad X + Z_Y \xrightarrow{ q_y } U + Z_Y
    &\qquad \qquad \text{(zealots)} \\
    U + Z_X \xrightarrow{ q_x } X + Z_X
    &\qquad \qquad U + Z_Y \xrightarrow{ q_y } Y + Z_Y
    &\qquad \qquad \text{(zealots)}. \\
  \end{array}
  \label{eq:chemical:ci}
\end{equation}
The cross-inhibition model \cite{Reina:PLOSONE:2015, Reina:PRE:2017} is a variation of the weighted voter model that includes a null state ($U$) on indecision as a necessary transition step between any other state \cite{Marvel2012}. We consider two alternative types of noise. The noise type 1 follows the transition schema of the model, by letting the individual go through the state $U$ before adopting a new opinion (see dashed lines in Fig.~\ref{fig:motifs}b).
To conduct a more complete comparison of the two models, we also consider noise type~2 which corresponds to the way noise is implemented in the voter model (see dotted lines in Fig.~\ref{fig:motifs}b).

\subsection*{Ordinary differential equations}
The change in the size of the populations committed to $X$ and $Y$ can be studied through a system of ODEs. We indicate with $S_X$, $S_Y$, $S_U$, $S_{Z_X}$, and $S_{Z_Y}$ the sizes of the populations in the state indicated in the subscript. As we only consider a symmetric number of zealots, we have $S_{Z_X} = S_{Z_Y}= S_Z$. For a total population of $S$ individuals (i.e.  $S=S_X + S_Y + S_U + 2S_{Z}$), we represent the proportions of individuals in each opinion state with $x=S_X/S$, $y=S_y/S$, $u=S_U/S$, and $z=S_Z/S$ (note that the total proportion of zealots in the population is $2z$). We consider only one type of asocial behaviour at a time. Therefore we first introduce a generic ODE system with the term $\alpha$ indicating the asocial behaviour that can be then substituted with either noise or zealotry.
Assuming a large population, for $S \rightarrow \infty$, the time
derivative of $x$ for the voter model is
\begin{equation}
  \label{eq:ode:vm:generic}
  \od{x}{t}= x y (q_x  - q_y) + \alpha \,.
\end{equation}
When we consider asocial behaviour in the form of noise, $\alpha = \sigma ( y - x)$, instead, when we consider zealot behaviour we have $\alpha = z ( q_x y - q_y x)$. The two alternative systems are
\begin{flalign}
 &\text{weighted voter model with symmetric noise $\sigma$} &
 \od{x}{t}= x y (q_x  - q_y) + \sigma ( y - x)\,; \label{eq:ode:vm:noise}\\
 &\text{weighted voter model with symmetric zealots $z$} &
 \od{x}{t}= q_x y (x  + z) - q_y x (y + z)\,. \label{eq:ode:vm:zealots}
\end{flalign}
In agreement with previous research \cite{Khalil2018}, it is straightforward to see how independent behaviour (noise) and zealots can be mathematically equivalent, that is when $\sigma ( y - x) = z ( q_x y - q_y x)$. The equivalence is present when either the options have equal quality, $q=q_x/q_y=1$, or when there are different rates for noise in the two populations, i.e. $\sigma_x \neq \sigma_y$. In our study, we keep the two models distinct in two ways. While the number of zealots is symmetrically distributed among the two alternatives, zealots recruit peers with rates proportional to their option's quality, instead noise is an asocial component that has a symmetric contribution for both options. Additionally, zealots are included as part of the total population $S$, therefore, increasing the number of zealots indirectly reduces the number of flexible individuals (that can change opinion), as $S_X+S_Y=S-2S_Z$; instead, increasing noise has no effect on the total number of to flexible individuals $S_X+S_Y=S$ (e.g., noise $\sigma$ can be the frequency of self-sourcing information). We find it more appropriate to consider zealots as part of the total population $S$, rather than an external source of influence.
Considering different models helps in showing that the results of our
analysis are consistent across different types of symmetric asocial
behaviour.

The stability analysis of Eqs.~\eqref{eq:ode:vm:noise} and \eqref{eq:ode:vm:zealots} can be simplified by replacing the proportion $y$ with $y=1-x-2z$, and, in the latter, by diving both sides by $q_y$ (note that the latter operation results in scaling the time by the quantity $q_y$, that is $\tau=q_y t$).  We define the quality ratio $q=q_x/q_y$. In this way, Eq.~\eqref{eq:ode:vm:zealots} can be rewritten as
\begin{equation}
  \label{eq:ode:vm:zealots:full}
  \od{x}{\tau}= q (1-x-2z) (x  + z) - x (1-x-z)\,.
\end{equation}

The ODE system for the cross-inhibition model with generic asocial behaviour reads as
\begin{equation}
  \label{eq:ode:ci:generic}
  \begin{cases}
  \od{x}{\tau}= x (q u -y ) + \alpha_x \\
  \od{y}{\tau}= y ( u - qx ) + \alpha_y \,.
  \end{cases}
\end{equation}
where $u=1-x-y-2z$.
When we instantiate asocial behaviour in the form of zealot behaviour,
$\alpha_x = z( qu - x)$ and $\alpha_y = z( u - qy )$, and we obtain
\begin{equation}
  \label{eq:ode:ci:zealots}
  \begin{cases}
  \od{x}{\tau}= q u (x + z) - x (y + z) \\
  \od{y}{\tau}= u  (y + z) - q y (x + z) \,,
  \end{cases}
\end{equation}
The cross-inhibition models with the two types of noise are presented in the Supplementary Note~1.

\paragraph{Stability analysis.} The voter model in both conditions---with noise or zealots---has two fixed points and only one of them is stable and in the positive plane $x\in[0,1]$. In the case of the voter model with noise (i.e.  $\sigma > 0$ and $z=0$), the stable fixed point is \begin{equation}
  \label{eq:vm:fp:noise}
  x^* =
  \frac{1}{2} + \frac{\sqrt{(q_x- q_y)^2+4 \sigma ^2}-2 \sigma }{2 (q_x- q_y)}
  \,.
\end{equation}
In the case of the voter model with zealots (i.e. $\sigma = 0$ and
$z > 0$), the stable fixed point is
\begin{equation}
  \label{eq:vm:fp:zealots}
  x^* =
  \frac{1}{2} +
  \frac{\sqrt{(q-1)^2 +z [ (q+1)^2 z -2 (q-1)^2 ]} -z(3q -1)}{2
  (q-1)}
\,.
\end{equation}
Note that Eqs.~(\ref{eq:vm:fp:noise}) and (\ref{eq:vm:fp:zealots}) only exist for $q \neq 1$ (that is, $q_x \neq q_y$), and instead for $q = 1$, in both conditions, the ODE system remains in decision deadlock with $x=y=0.5$.
The analysis of the fixed point for $q \neq 1$ reveals that for relatively low values of noise, or zealotry, the system can reach a large majority only when the quality difference is large, i.e., $q \gg 1$ or symmetrically $q \approx 0$. Figures~\ref{fig:noise}a-d,~\ref{fig:qratio}a,b,~and~\ref{fig:bifurcation}c show the predicted agreement levels, according to Eqs.~(\ref{eq:vm:fp:noise}) and (\ref{eq:vm:fp:zealots}) for various values of the quality ratio $q$, the noise level $\sigma$, and proportion of zealots $z$.

The cross-inhibition model has four fixed points that change their stability as a function of the system's parameters. For the symmetric quality case, $q=1$, with zealots $z\ge0$, the fixed points (pre-bifurcation) are \begin{equation}
  \label{eq:ci:fp:zealots}
  x^*=
  \frac{1}{2} \left(1-3 z \pm \sqrt{5 z^2-6 z+1}\right) \,;\qquad
  y^*=
  \frac{1}{2} \left(1-3 z \mp \sqrt{5 z^2-6 z+1}\right)
\,.
\end{equation}
For the asymmetric case, $q\neq 1$, the results are more complex.  While it is possible to derive the fixed points' mathematical equations and their stability conditions through standard software (e.g., see the Mathematica notebook provided in the Supplementary Software), the complexity of the equations makes it impossible to study them from their mathematical form. Instead, we can plot the dynamics for representative cases and so interpret the general dynamics of the system. In particular, in Figure~\ref{fig:bifurcation}c we display the fixed point of population $x$ for both the cross-inhibition and the voter models, so that we can appreciate the differences between the two. It is noteworthy that (except for extremely high values of $z$), the cross-inhibition model always grants a larger majority in favour of the best option than the voter model. However, as shown in Figures~\ref{fig:qratio}c,d,~and~\ref{fig:bifurcation}b, the cross-inhibition model can have more than one stable state, leading to a consensus for the inferior alternative. Counterintuitively, the lower branch, in favour of the inferior alternative, is only present for low values of asocial behaviour and vanishes as the number of zealots increases. Therefore, it appears that zealotry can improve the accuracy of the cross-inhibition model by removing the possibility of selecting the inferior option. The minimum quantity of zealot behaviour $z$ to make the lower branch vanish decreases as the quality difference increases. This effect can be measured by plotting the point of the fold bifurcation as a function of the quality ratio $q$; Figure~\ref{fig:bifurcation}b shows that the parameter space where the lower branch is present reaches the maximum $z=0.2$ for $q=1$ and rapidly decreases as $q$ gets larger than 1.

\subsection*{Master equations}

The dynamics of finite-sized systems can be studied through the master equations. Here, for both models, we present the master equations and find their analytical solution at equilibrium (that is, the stationary probability distribution, SPD), finally, we illustrate the numerical analysis employed to study the transitory temporal dynamics. 

\paragraph{Master equation of the weighted voter model.} We first define the rates at which transitions occur: $T^{+x}_{x=k}$ is the rate by which the population committed to $X$, with size $S_X=k$, increases of one individual, and $T^{-x}_{x=k}$ is the rate by which it decreases by one individual. We recall that the total population has size $S$ which is composed of subpopulations in different states: either committed to options $X$ or $Y$ with sizes $S_X$ and $S_Y$, or in state zealot with sizes $S_{Z_X}$ and $S_{Z_Y}$, that in our study we consider symmetric $S_{Z_X}=S_{Z_Y}=S_Z$. Therefore, the system is completely specified (fully defined)
computing only the changes for the population committed to $X$ because $S_Y=S-S_X-2S_Z$.
In the voter model,
the rate $T^{+x}_{x=k}$ describes when an individual in state $Y$
directly switches to $X$ after an interaction with an individual
committed to $X$ (either a susceptible individual in state $X$ or a
zealot in state $Z_X$), that happens with frequency proportional to
option $X$'s quality $q_x$. Otherwise, when there are no zealots
$S_Z=0$, the transition can also be caused by asocial noise
$\sigma>0$.  Therefore,
\begin{equation}
  \begin{array}{llll}
    \text{with noise:}&
     T_{x=k}^{+x}=(S-k) \left(q_x\frac{k}{S-1} + \sigma \right);&
                                                                       \qquad
     \text{with zealots:}&
     T_{x=k}^{+x}= (S-k-2S_Z) \, q_x \left(\frac{k+S_Z}{S-1}\right).\\
  \end{array}
  \label{eq:master:vm:t+}
\end{equation}
Symmetrically, the population committed to $X$ decreases by one when one of its individual (state $X$) directly switches to $Y$ after an interaction with a susceptible individual in state $Y$ or a zealot in state $Z_Y$ (with frequency proportional to option $Y$'s quality $q_y$), or alternatively through asocial noise $\sigma>0$. The reduction rate $T^{-x}_{x=k}$ is
\begin{equation}
  \begin{array}{llll}
    \text{with noise:}&
     T_{x=k}^{-x}=k \left(q_y \frac{S-k}{S-1} + \sigma \right);&
                                                                       \qquad
     \text{with zealots:}&
     T_{x=k}^{-x}= k \, q_y \left(\frac{S-k-2S_Z+S_Z}{S-1}\right).\\
  \end{array}
  \label{eq:master:vm:t-}
\end{equation}

The system state is characterised by the probability $P_{x=k}(t)$ that the number of individuals in state $X$ at time $t$ are $S_X=k$. Through the rates of Eqs.~\eqref{eq:master:vm:t+} and \eqref{eq:master:vm:t-}, we can define the master equation of the weighted voter model which
describes how the probability $P_{x=k}(t)$ changes over time:
\begin{equation}
\od{P_{x=k}(t)}{t}=T^{+x}_{x=k-1} P^{}_{x=k-1}(t) + T^{-x}_{x=k+1} P^{}_{x=k+1}(t)
-  (T_{x=k}^{+x} + T_{x=k}^{-x})P^{}_{x=k}(t).
\label{eq:master:vm}
\end{equation}
The first two terms of the rhs of Eq.~\eqref{eq:master:vm} account for processes in which the number of individuals in state $X$ after the event equals to $S_X=k$, while the last term accounts for the complementary loss processes where $S_X$ increases or decreases of one unit (i.e. $(S_X=k) \rightarrow (S_X=k+1)$ or $(S_X=k) \rightarrow (S_X=k-1)$, respectively).

\paragraph{Stationary probability distribution of the weighted voter
  model.} In stationary conditions, the master equation reaches an equilibrium, that is $\od{P_{x=k}(t)}{t}=0$, and therefore the probability distribution is stable over time. The stationary master equation obtained from Eq.~\eqref{eq:master:vm} is
\begin{equation}
\label{eq:master:stationary}
T^{+x}_{x=k-1} P^*_{x=k-1} + T^{-x}_{x=k+1} P^*_{x=k+1}
-(T_{x=k}^{+x} + T_{x=k}^{-x})P^*_{x=k}=0 \,,
\end{equation}
where $P^*_{x=k}=\text{lim}_{t\rightarrow \infty} P_{x=k}(t)$ indicates the stationary solution that is independent of time $t$.
As the process can be considered time reversible at equilibrium conditions, the SPD can be derived from the detailed balance principle, which yields \begin{equation}
  T^{+x}_{x=k-1} P^*_{x=k-1}=T^{-x}_{x=k} P^*_{x=k} \,.
  \label{eq:balance:vm}
\end{equation}
By iterating Eq.\eqref{eq:balance:vm}, we obtain the SPD:
\begin{equation}
\label{eq:spd:vm}
P^*_{x=k}=P^*_{x=0}
\prod_{j=0}^{k-1}\frac{T_{x=j}^{+x}}{T_{x=j+1}^{-x}} \,,
\end{equation}
where the normalization condition $\sum_{k=0}^{S-2S_Z} P^*_{x=k}=1$ allows us to
compute $P^*_{x=0}$ as:
\begin{equation}
\label{eq:spd:vm:P0}
  P^*_{x=0}=\frac{1}{1+\sum_{k=1}^{S-2S_Z} \prod_{j=0}^{k-1}\frac{T_{x=j}^{+x}}{T_{x=j+1}^{-x}}}\,.
\end{equation}
Note that the summation on the denominator goes to a maximum of $S-2S_Z$, which is the maximum size that the population $S_X$ can be.
It is also obvious that the complementary probability $P_{y=k}^*$ for the population committed to option $Y$ can be computed as $P_{y=k}^*=1 - P_{x=S-k-2S_Z}^*$.
The expanded forms of Eq.~\eqref{eq:spd:vm} for both asocial mechanisms (noise and zealotry), with the rates from Eqs.~\eqref{eq:master:vm:t+} and~\eqref{eq:master:vm:t-}, are presented in the Supplementary Note~2.

The SPD of the difference of the populations committed to the two options, illustrated in Fig.~\ref{fig:motifs}e, \ref{fig:noise}a-d, and ~\ref{fig:qratio}a-b, is computed as \begin{equation} \label{eq:spd:vm:diff} P_{x-y=k}^*=P_{x=\frac{S-2S_Z-k}{2}}^* \,.  \end{equation}

\paragraph{Master equation of the cross-inhibition model.}  For the cross-inhibition model, the increment and reduction rates necessary to define the master equation are $T^{+x}_{x=a,y=b}$, $T^{-x}_{x=a,y=b}$, $T^{+y}_{x=a,y=b}$, and $T^{-y}_{x=a,y=b}$. These rates are increment (superscript $+x$ or $+y$) or reductions (superscript $-x$ or $-y$) of one individual in the population committed to $X$ ($+x$ or $-x$) or to $Y$ ($+y$ or $-y$) given that the committed populations are $S_X=a$ and $S_Y=b$.   Here, we only describe rates for zealots and noise of type 1 (see Eq.~\eqref{eq:chemical:ci}), because the SPD of the master equation with noise type 2 cannot be computed analytically using the detailed balance principle due to transitions happening both from state $U$ to states $X$ and $Y$, and directly between states $X$ and $Y$. Therefore, we only conducted numerical analysis for the cross-inhibition model with noise type 2 (see Section on the stochastic simulation algorithm). For the other two types of asocial behaviours, increments by one individual in the committed populations, $S_X$ or $S_Y$, occur when individuals of that population, or zealots for that option, recruit an uncommitted individual; otherwise, through asocial noise. Rates $T^{+x}_{x=a,y=b}$ and $T^{+y}_{x=a,y=b}$ are \begin{equation}
  \begin{array}{lll}
    \text{with noise type 1:}&
  T^{+x}_{x=a,y=b}= (S-a-b) \left(q_x \frac{a}{S-1} + \sigma\right);&
  
  T^{+y}_{x=a,y=b}= (S-a-b) \left(q_y \frac{b}{S-1} + \sigma\right)\\
 \text{with zealots:}&
  T^{+x}_{x=a,y=b}=q_x (S-a-b-2S_Z) \left(\frac{a+S_Z}{S-1}\right);&
  
  T^{+y}_{x=a,y=b}=q_y (S-a-b-2S_Z) \left(\frac{b+S_Z}{S-1}\right).\\
  \end{array}
  \label{eq:master:ci:t+}
\end{equation}
Instead, reduction by one individual from populations committed to $X$
or $Y$ occurs when individuals committed to different options interact
with each other, or spontaneously, through asocial noise. Rates
$T^{-x}_{x=a,y=b}$ and $T^{-y}_{x=a,y=b}$ are
\begin{equation}
  \begin{array}{lll}
    \text{with noise type 1:}&
  T^{-x}_{x=a,y=b}= a \left(q_y \frac{b}{S-1} + \sigma\right);&
  
  T^{-y}_{x=a,y=b}= b \left(q_x \frac{a}{S-1} + \sigma\right)\\
 \text{with zealots:}&
  T^{-x}_{x=a,y=b}=a\,q_y  \left(\frac{b+S_Z}{S-1}\right);&
  
  T^{-y}_{x=a,y=b}=b\,q_x  \left(\frac{a+S_Z}{S-1}\right).\\
  \end{array}
  \label{eq:master:ci:t-}
\end{equation}

The master equation for the cross-inhibition model describes the
change over time of the probability $P_{x=a,y=b}(t)$ which indicates
that number of individual committed to option $X$ and $Y$ at time $t$
are $S_X=a$ and $S_X=b$, respectively. Because each individual (except
for the zealots that do not change state) can be in three possible
states---committed to $X$, committed to $Y$, or uncommitted (state
$U$)---we have $S_U=S - S_X-S_Y-2S_Z$ and considering the sizes $S_X$
and $S_Y$ is sufficient to completely specify (i.e. to fully define) the system. The master equation is
\begin{equation}
  \label{eq:master:ci}
  \begin{array}{lll}
  \od{P_{x=a,y=b}(t)}{t}&=& T^{+x}_{x=a-1,y=b} P_{x=a-1,y=b}(t) +
                           T^{-x}_{x=a+1,y=b} P_{x=a+1,y=b}(t) \\
    &+&T^{+y}_{x=a,y=b-1} P_{x=a,y=b-1}(t) +
        T^{-y}_{x=a,y=b+1} P_{x=a,y=b+1}(t) \\
        &-& ( T^{+x}_{x=a,y=b}  + T^{-x}_{x=a,y=b} 
            + T^{+y}_{x=a,y=b}  + T^{-y}_{x=a,y=b} )
            P_{x=a,y=b}(t) \\
  \end{array}
\end{equation}

\paragraph{Stationary probability distribution of the cross-inhibition
  model.}
We compute the stationary solution
$P_{x=a,y=b}^*=\text{lim}_{t\rightarrow \infty} P_{x=a,y=b}(t)$ of the
master equation when the dynamics have converged
($\od{P_{x=a,y=b}(t)}{t}=0$) and the probability distribution is
independent of time. Because there are no direct transitions from
state $X$ to state $Y$ and viceversa, but the agents always pass
through the uncommitted state $U$, each transition consists in a
change of one individual (increase or reduction) either in $A$ or in
$B$ and it can never change both populations at once.  Therefore, we
can employ the detailed balance principle and obtain
\begin{flalign}
  T^{+x}_{x=a-1,y=b} P^*_{x=a-1,y=b}=T^{-x}_{x=a,y=b} P^*_{x=a,y=b}
  \,,&
  \quad \text{and}  \label{eq:balance:ci1}\\
  T^{+y}_{x=a,y=b-1} P^*_{x=a,y=b-1}=T^{-y}_{x=a,y=b} P^*_{x=a,y=b}
  \,.&
  \label{eq:balance:ci2}
\end{flalign}
By iterating Eqs.~\eqref{eq:balance:ci1} and~\eqref{eq:balance:ci2},
we can compute, respectively, 
\begin{flalign}
P^*_{x=a,y=0}=P^*_{x=0,y=0}
\prod_{j=0}^{a-1}\frac{T_{x=j,y=0}^{+x}}{T_{x=j+1,y=0}^{-x}}\,, &
\quad \text{and}
\label{eq:spd:ci:y0}\\
P^*_{x=0,y=b}=P^*_{x=0,y=0}
\prod_{j=0}^{b-1}\frac{T_{x=0,y=j}^{+y}}{T_{x=0,y=j+1}^{-y}}\,.&
\label{eq:spd:ci:x0}
\end{flalign}
Through Eq.~\eqref{eq:spd:ci:y0} and iterating
Eq.~\eqref{eq:balance:ci2}, we can compute the generic SPD equation:
\begin{equation}
P^*_{x=a,y=b}=P^*_{x=a,y=0}
  \prod_{j=0}^{b-1}\frac{T_{x=a,y=j}^{+y}}{T_{x=a,y=j+1}^{-y}}
\label{eq:spd:ci}
\end{equation}
Finally, the value of $P^*_{x=0,y=0}$ is computed through the
normalisation condition
\begin{equation*}
\sum_{a=0}^{(S-2S_Z)} \sum_{b=0}^{(S-2S_Z-a)}
P^*_{x=a,y=b}=1 \,,
\end{equation*}
which can be written as
\begin{equation}
P^*_{x=0,y=0}+\sum_{b=1}^{S-2S_Z}
P^*_{x=0,y=b}+
\sum_{a=1}^{S-2S_Z} \left( P^*_{x=a,y=0} + \sum_{b=1}^{(S-2S_Z-a)} P^*_{x=a,y=b} \right)=1\,.
  \label{eq:spd:ci:normalisation}
\end{equation}
Replacing Eqs.~\eqref{eq:spd:ci:y0}, \eqref{eq:spd:ci:x0},
and~\eqref{eq:spd:ci} in Eq.~\eqref{eq:spd:ci:normalisation}, we obtain
\begin{equation}
  P^*_{x=0,y=0}=\left[ 1 + \sum_{b=1}^{(S-2S_Z)}
\prod_{j=0}^{b-1}\frac{T_{x=0,y=j}^{+y}}{T_{x=0,y=j+1}^{-y}} +
\sum_{a=1}^{S-2S_Z} \prod_{j=0}^{a-1}\frac{T_{x=j,y=0}^{+x}}{T_{x=j+1,y=0}^{-x}}\left( 1+ \sum_{b=1}^{(S-2S_Z-a)} \prod_{k=0}^{b-1}
  \frac{T_{x=a,y=k}^{+y}}{T_{x=a,y=k+1}^{-y}}\right) \right]^{-1} .
  \label{eq:spd:ci:P00}
\end{equation}

The expanded forms of Eq.~\eqref{eq:spd:ci} for both asocial mechanisms (noise and zealotry), with the rates from Eqs.~\eqref{eq:master:ci:t+} and~\eqref{eq:master:ci:t-}, are presented in the Supplementary Note~2.

The SPD for a single state
is computed as the sum of the SPD for all values of the other state, that is,
\begin{equation}
\label{eq:spd:ci:single}
P_{x=a}^*=\sum_{b=0}^{S-2S_Z-a}P_{x=a,y=b}^* \,, \qquad\text{and}\qquad
P_{y=b}^*=\sum_{a=0}^{S-2S_Z-b}P_{x=a,y=b}^* \,.
\end{equation}
Instead, when we display the SPD of the difference of the populations committed to the two options, as for instance in Figs.~\ref{fig:motifs}f and~\ref{fig:qratio}c-d, we compute the following quantity
\begin{equation}
\label{eq:spd:ci:diff}
P_{x-y=k}^*=\sum_{(a,b) \in D_k } P_{x=a,y=b}^* \,, \quad\text{with}\quad
D_k = \{ (a,b) \,|\, a,b \in \mathbb{N}^0 \,\&\, a+b \le S-2S_Z \,\&\, a-b = k \}  \,.
\end{equation}

\paragraph{Stochastic simulation algorithm.} We complemented the analytical form of the stationary probability distribution with its numerical approximation, which we computed through Gillespie's stochastic simulation algorithm \cite{Gillespie2013}. Supplementary Figure~S2 shows that analytical and numerical solutions have a precise match for both the voter and the cross-inhibition model with both asocial mechanisms (noise and zealotry). As indicated earlier, we could not find the analytical solution of the SPD for the cross-inhibition model with noise type 2 because there are transitions between all three states and employing the detailed balance principle here becomes unpractical. Therefore, the results of Figures~\ref{fig:noise}g-h, which show the effect of noise type 2, are computed through the Gillespie algorithm ($10^4$ runs per condition). For every run, we initialise the system with a random initial condition and store the amount of time spent in each state throughout long runs ($10^5$ time-units, with transition rate speed defined as from Eqs.~\eqref{eq:chemical:vm} and~\eqref{eq:chemical:ci}). The time in each state is normalised by the total simulation length so that it approximates the stationary probability distribution \cite{Gillespie2013}.

\paragraph{Mean switching time.} We compute the mean switching time (MST) as the expected time to move from a state of full consensus in favour of $Y$---the option with lower quality---to a state of stable consensus for $X$, the option with superior quality. The MST
illustrated in Fig.~\ref{fig:bifurcation}a is the average of 500 runs of Gillespie's stochastic simulation algorithm, running one million time units. A switch towards consensus for the superior option $X$ happens when the subpopulation in favour of $X$ has size $S_X >= \hat{S}_{X}$ for more than 10 time units. We do not include these 10 time units in the MST. The term $\hat{S}_{X}$ is the minimum size of the subpopulation committed to $X$ to consider the system in a state of consensus for $X$. The size threshold $\hat{S}_{X}$ is calculated numerically from a very long run of the SSA ($10^6$ time-units) initialised at $S_X = S-2S_Z$ and $S_Y= 0$. While the population remains in the state of consensus for $X$, the actual size of the subpopulation $S_X$ fluctuates over time. Assuming a normal distribution of the fluctuations, we compute $\hat{S}_{X}$ as the average of $S_X$ when $S_X > (S-2S_Z)/2$,  subtracted by 3 times the standard deviation of the considered data points.

\subsection*{Swarm robotics experiments}

We ran four experiments with a swarm of 100 Kilobots \cite{Rubenstein2014}, which are low-cost small-sized robots that can move on a flat surface, display their internal state through a coloured LED light, and broadcast 9-byte infrared messages to neighbours in a range of \unit[10]{cm} (see Figure \ref{fig:motifs}d). Robot's movements are limited to (approximately) straight motion at a speed of about \unitfrac[1]{cm}{s} and rotation in place at about \unitfrac[40]{$^\circ$}{s}. In our experiments, by alternating straight motion and rotations, the robots diffused throughout the \unit[$1\times1$]{m$^2$} environment, using the same mobility model of \cite{Talamali:SciRobot:2021}, and hence changed their neighbourhood over time. Every 30 seconds, the robot read the last message received from its neighbours, that used to update its opinion through one of the two state machines of Figure~\ref{fig:motifs}.
As shown in previous research \cite{Reina:DARS:2016,Reina:SwInt:2015}, using a relatively low frequency of robot's opinion update (\unit[30]{s}) compared with the random diffusion speed (\unitfrac[1]{cm}{s}) allows obtaining a qualitative good agreement between the macroscopic models---which assume a well-mixed interaction topology---and the robotic implementation---which relies on local interactions in a range of \unit[10]{cm}.
The robots showed their opinion via the coloured LED (option X as red, option Y as blue, and no opinion as green). The experiments were recorded using an overhead camera (the videos are available as Supplementary Movies~1-5), and the robots' position and state---that is, their light's colour---were tracked through the ARK system \cite{Reina:RAL:2017}. In every experiment, we included a number of zealot robots, which ran the same algorithm of the others except for refraining to update their opinion. We began each experiment with the swarm opinion equally split, 50-50, between the two options. Each experiment lasted 60 minutes and the algorithms ran by the robots are open-source and available in the Supplementary Software.


\subsection*{Data availability}
Videos of the robot experiments are available as Supplementary Movies~1-5. Supplementary Movie~1 is also available at \url{https://youtu.be/mQtLhMqdVWg}.

\subsection*{Code availability}
All simulation code, Mathematica notebooks, and robot code to reproduce the analyses and experiments presented herein are
available as Supplementary Software in the GitHub repository \url{https://github.com/rainazakir/asocialbehaviour}.

\subsection*{Author Contributions}
A.R. conceived the original idea and directed the project. A.R. performed the mean-field ODE analysis. A.R., R.Z., G.D.M., and E.F. derived the solution of the master equation. R.Z. generated the results from the master equations. A.R. and R.Z. generated the figures. A.R. designed and implemented the robot control code and conducted the robot experiments. A.R. recorded and edited the videos. A.R. wrote the first draft of the manuscript and all authors edited the manuscript.

\subsection*{Competing interests}
The authors declare that they have no competing interests.

\subsection*{Acknowledgements}
The authors thank Alex J. Cope for his help in taking the photo of Figure 1d. A.R. and R.Z. acknowledge support from the Belgian F.R.S.-FNRS, of which they are Charg\'{e} de Recherches and FRIA Doctoral Student, respectively.

\bibliographystyle{unsrt}
\bibliography{refs_zealots}

\begin{thebibliography}{10}

\bibitem{Castellano2009}
Claudio Castellano, Santo Fortunato, and Vittorio Loreto.
\newblock Statistical physics of social dynamics.
\newblock {\em Reviews of Modern Physics}, 81(2):591--646, 2009.

\bibitem{Conradt2009}
Larissa Conradt and Christian List.
\newblock Group decisions in humans and animals: a survey.
\newblock {\em Philosophical Transactions of the Royal Society B: Biological
  Sciences}, 364(1518):719--742, 2009.

\bibitem{Baronchelli2018}
Andrea Baronchelli.
\newblock The emergence of consensus: a primer.
\newblock {\em Royal Society Open Science}, 5(2):172189, 2018.

\bibitem{seeley2011}
Thomas Seeley.
\newblock {\em Honeybee Democracy}.
\newblock Princeton University Press, 2011.

\bibitem{Valentini2017}
Gabriele Valentini.
\newblock {\em Achieving Consensus in Robot Swarms: {D}esign and Analysis of
  Strategies for the best-of-$n$ Problem}, volume 706 of {\em Studies in
  Computational Intelligence}.
\newblock Springer International Publishing, Cham, Switzerland, 2017.

\bibitem{Reina:SwInt:2021}
Andreagiovanni Reina, Eliseo Ferrante, and Gabriele Valentini.
\newblock Collective decision-making in living and artificial systems:
  editorial.
\newblock {\em Swarm Intelligence}, 15(1-2):1--6, 2021.

\bibitem{Clifford1973}
Peter Clifford and Aidan Sudbury.
\newblock A model for spatial conflict.
\newblock {\em Biometrika}, 60(3):581--588, 1973.

\bibitem{Holley1975}
Richard~A Holley and Thomas~Milton Liggett.
\newblock Ergodic theorems for weakly interacting infinite systems and the
  voter model.
\newblock {\em The annals of probability}, 3(4):643--663, 1975.

\bibitem{Jhawar2020}
Jitesh Jhawar, Richard~G. Morris, U.~R. Amith-Kumar, M.~{Danny Raj}, Tim
  Rogers, Harikrishnan Rajendran, and Vishwesha Guttal.
\newblock Noise-induced schooling of fish.
\newblock {\em Nature Physics}, 16(4):488--493, 2020.

\bibitem{FernandezGracia2014}
Juan Fern{\'{a}}ndez-Gracia, Krzysztof Suchecki, Jos{\'{e}}~J. Ramasco, Maxi
  {San Miguel}, and V{\'{i}}ctor~M. Egu{\'{i}}luz.
\newblock Is the voter model a model for voters?
\newblock {\em Physical Review Letters}, 112(15):158701, 2014.

\bibitem{Zillio2005}
Tommaso Zillio, Igor Volkov, Jayanth~R. Banavar, Stephen~P. Hubbell, and Amos
  Maritan.
\newblock Spatial scaling in model plant communities.
\newblock {\em Physical Review Letters}, 95(9):098101, 2005.

\bibitem{Redner2019}
Sidney Redner.
\newblock Reality-inspired voter models: A mini-review.
\newblock {\em Comptes Rendus Physique}, 20(4):275--292, 2019.

\bibitem{Mobilia2007}
Mauro Mobilia, A.~Petersen, and Sidney Redner.
\newblock On the role of zealotry in the voter model.
\newblock {\em Journal of Statistical Mechanics: Theory and Experiment},
  (08):P08029--P08029, 2007.

\bibitem{Khalil2018}
Nagi Khalil, Maxi {San Miguel}, and Raul Toral.
\newblock Zealots in the mean-field noisy voter model.
\newblock {\em Physical Review E}, 97(1):012310, 2018.

\bibitem{Galam2007}
Serge Galam and Frans Jacobs.
\newblock The role of inflexible minorities in the breaking of democratic
  opinion dynamics.
\newblock {\em Physica A: Statistical Mechanics and its Applications},
  381:366--376, 2007.

\bibitem{Seeley2012}
Thomas~D. Seeley, P.~Kirk Visscher, Thomas Schlegel, Patrick~M. Hogan, Nigel~R.
  Franks, and James A.~R. Marshall.
\newblock Stop signals provide cross inhibition in collective decision-making
  by honeybee swarms.
\newblock {\em Science}, 335(6064):108--111, 2012.

\bibitem{Reina:PLOSONE:2015}
Andreagiovanni Reina, Gabriele Valentini, Cristian Fern\'{a}ndez-Oto, Marco
  Dorigo, and Vito Trianni.
\newblock A design pattern for decentralised decision making.
\newblock {\em PLoS ONE}, 10(10):e0140950, 2015.

\bibitem{Reina:PRE:2017}
Andreagiovanni Reina, James A.~R. Marshall, Vito Trianni, and Thomas Bose.
\newblock Model of the best-of-{N} nest-site selection process in honeybees.
\newblock {\em Physical Review E}, 95(5):052411, 2017.

\bibitem{Bogacz2006}
Rafal Bogacz, Eric Brown, Jeff Moehlis, Philip Holmes, and Jonathan~D Cohen.
\newblock The physics of optimal decision making: A formal analysis of models
  of performance in two-alternative forced-choice tasks.
\newblock {\em Psychological Review}, 113(4):700--765, 2006.

\bibitem{Higgins2009}
Fiona Higgins, Allan Tomlinson, and Keith~M. Martin.
\newblock Threats to the swarm: Security considerations for swarm robotics.
\newblock {\em International Journal on Advances in Security}, 2(2-3):288--297,
  2009.

\bibitem{DeMasi2021}
Giulia De~Masi, Judhi Prasetyo, Raina Zakir, Nikita Mankovskii, Eliseo
  Ferrante, and Elio Tuci.
\newblock Robot swarm democracy: the importance of informed individuals against
  zealots.
\newblock {\em Swarm Intelligence}, 15(4):315--338, 2021.

\bibitem{Rubenstein2014}
Michael Rubenstein, Christian Ahler, Nick Hoff, Adrian Cabrera, and Radhika
  Nagpal.
\newblock Kilobot: A low cost robot with scalable operations designed for
  collective behaviors.
\newblock {\em Robotics and Autonomous Systems}, 62(7):966--975, 2014.

\bibitem{Galam2011}
Serge Galam.
\newblock Collective beliefs versus individual inflexibility: The unavoidable
  biases of a public debate.
\newblock {\em Physica A: Statistical Mechanics and its Applications},
  390(17):3036--3054, 2011.

\bibitem{Galam2020}
Serge Galam and Taksu Cheon.
\newblock Tipping points in opinion dynamics: A universal formula in five
  dimensions.
\newblock {\em Frontiers in Physics}, 8, 2020.

\bibitem{Zakir:ANTS:2022}
Raina Zakir, Marco Dorigo, and Andreagiovanni Reina.
\newblock Robot swarms break decision deadlocks in collective perception
  through cross-inhibition.
\newblock In {M. Dorigo et al.}, editor, {\em Swarm Intelligence (ANTS 2022)},
  volume 13491 of {\em LNCS}, pages 209--221. Springer, Cham, 2022.

\bibitem{holldobler1990ants}
Bert H{\"o}lldobler and Edward~O Wilson.
\newblock {\em The ants}.
\newblock Harvard University Press, 1990.

\bibitem{Cardelli2017}
Luca Cardelli, Rosa~D. Hernansaiz-Ballesteros, Neil Dalchau, and Attila
  Csik{\'{a}}sz-Nagy.
\newblock Efficient switches in biology and computer science.
\newblock {\em PLOS Computational Biology}, 13(1):e1005100, 2017.

\bibitem{Franci:SwInt:2021}
Alessio Franci, Anastasia Bizyaeva, Shinkyu Park, and Naomi~E. Leonard.
\newblock Analysis and control of agreement and disagreement opinion cascades.
\newblock {\em Swarm Intelligence}, 15(1--2):47--82, 2021.

\bibitem{Marshall2009}
James A.~R. Marshall, Rafal Bogacz, Anna Dornhaus, Robert Planqu{\'{e}}, Tim
  Kovacs, and Nigel~R. Franks.
\newblock On optimal decision-making in brains and social insect colonies.
\newblock {\em Journal of The Royal Society Interface}, 6(40):1065--1074, 2009.

\bibitem{Reina:scirep:2018}
Andreagiovanni Reina, Thomas Bose, Vito Trianni, and James A.~R. Marshall.
\newblock Psychophysical laws and the superorganism.
\newblock {\em Scientific Reports}, 8(1):4387, 2018.

\bibitem{Borofsky2020}
Talia Borofsky, Victor~J. Barranca, Rebecca Zhou, Dora von Trentini, Robert~L.
  Broadrup, and Christopher Mayack.
\newblock Hive minded: like neurons, honey bees collectively integrate negative
  feedback to regulate decisions.
\newblock {\em Animal Behaviour}, 168:33--44, 2020.

\bibitem{Biancalani2014}
Tommaso Biancalani, Louise Dyson, and Alan~J. McKane.
\newblock Noise-induced bistable states and their mean switching time in
  foraging colonies.
\newblock {\em Physical Review Letters}, 112(3):038101, 2014.

\bibitem{Herrerias-Azcue2019}
Francisco Herrer{\'{i}}as-Azcu{\'{e}} and Tobias Galla.
\newblock Consensus and diversity in multistate noisy voter models.
\newblock {\em Physical Review E}, 100(2):022304, 2019.

\bibitem{Khaluf2017}
Yara Khaluf, Carlo Pinciroli, Gabriele Valentini, and Heiko Hamann.
\newblock The impact of agent density on scalability in collective systems:
  noise-induced versus majority-based bistability.
\newblock {\em Swarm Intelligence}, 11(2):155--179, 2017.

\bibitem{Valentini2016}
Gabriele Valentini, Eliseo Ferrante, Heiko Hamann, and Marco Dorigo.
\newblock Collective decision with 100 kilobots: speed versus accuracy in
  binary discrimination problems.
\newblock {\em Autonomous Agents and Multi-Agent Systems}, 30(3):553--580,
  2016.

\bibitem{Reina:DARS:2016}
Andreagiovanni Reina, Thomas Bose, Vito Trianni, and James A.~R. Marshall.
\newblock Effects of spatiality on value-sensitive decisions made by robot
  swarms.
\newblock In {\em Distributed Autonomous Robotic Systems (DARS 2016): The 13th
  International Symposium}, volume~6 of {\em {SPAR}}, pages 461--473. Springer
  International Publishing, Cham, Switzerland, 2018.

\bibitem{Talamali:SciRobot:2021}
Mohamed~S. Talamali, Arindam Saha, James~A. R., and Andreagiovanni Reina.
\newblock When less is more: Robot swarms adapt better to changes with
  constrained communication.
\newblock {\em Science Robotics}, 6(56):eabf1416, 2021.

\bibitem{Valentini2014}
Gabriele Valentini, Heiko Hamann, and Marco Dorigo.
\newblock Self-organized collective decision making: The weighted voter model.
\newblock In {\em AAMAS '14: Proceedings of the 2014 international conference
  on Autonomous Agents and Multi-Agent Systems}, pages 45--52, 2014.

\bibitem{Bizyaeva2020}
Anastasia Bizyaeva, Alessio Franci, and Naomi~Ehrich Leonard.
\newblock Nonlinear opinion dynamics with tunable sensitivity.
\newblock {\em IEEE Transactions on Automatic Control}, 68(3):1415--1430, 2023.

\bibitem{Pirrone2014}
Angelo Pirrone, Tom Stafford, and James A.~R. Marshall.
\newblock When natural selection should optimize speed-accuracy trade-offs.
\newblock {\em Frontiers in Neuroscience}, 8(73):1--5, 2014.

\bibitem{Talamali:ICRA:2019}
Mohamed~S. Talamali, James A.~R. Marshall, Thomas Bose, and Andreagiovanni
  Reina.
\newblock Improving collective decision accuracy via time-varying
  cross-inhibition.
\newblock In {\em Proceedings of the 2019 IEEE International Conference on
  Robotics and Automation (ICRA 2019)}, pages 9652--9659. IEEE, 2019.

\bibitem{Bose:COBS:2017}
Thomas Bose, Andreagiovanni Reina, and James A.~R. Marshall.
\newblock Collective decision-making.
\newblock {\em Current Opinion in Behavioral Sciences}, 6:30--34, 2017.

\bibitem{Pirrone:TiCS:2021}
Angelo Pirrone, Andreagiovanni Reina, Tom Stafford, James A.~R. Marshall, and
  Fernand Gobet.
\newblock Magnitude-sensitivity: Rethinking decision-making.
\newblock {\em Trends in Cognitive Sciences}, in press, 2021.

\bibitem{Rajendran2022}
Harikrishnan Rajendran, Amir Haluts, Nir~S. Gov, and Ofer Feinerman.
\newblock Ants resort to majority concession to reach democratic consensus in
  the presence of a persistent minority.
\newblock {\em Current Biology}, 32(3):645--653.e8, 2022.

\bibitem{tajima2016optimal}
Satohiro Tajima, Jan Drugowitsch, and Alexandre Pouget.
\newblock Optimal policy for value-based decision-making.
\newblock {\em Nature Communications}, 7:12400, 2016.

\bibitem{Tsimring2014}
Lev~S. Tsimring.
\newblock Noise in biology.
\newblock {\em Reports on Progress in Physics}, 77(2):026601, 2014.

\bibitem{Schweitzer2018}
Frank Schweitzer.
\newblock Sociophysics.
\newblock {\em Physics Today}, 71(2):40--46, 2018.

\bibitem{Verma2014}
Gunjan Verma, Ananthram Swami, and Kevin Chan.
\newblock The impact of competing zealots on opinion dynamics.
\newblock {\em Physica A: Statistical Mechanics and its Applications},
  395:310--331, 2014.

\bibitem{Waagen2015}
Alex Waagen, Gunjan Verma, Kevin Chan, Ananthram Swami, and Raissa D'Souza.
\newblock Effect of zealotry in high-dimensional opinion dynamics models.
\newblock {\em Physical Review E}, 91(2):022811, 2015.

\bibitem{Shirado2017}
Hirokazu Shirado and Nicholas~A. Christakis.
\newblock Locally noisy autonomous agents improve global human coordination in
  network experiments.
\newblock {\em Nature}, 545(7654):370--374, 2017.

\bibitem{Rausch:SwInt:2019}
Ilja Rausch, Andreagiovanni Reina, Pieter Simoens, and Yara Khaluf.
\newblock Coherent collective behaviour emerging from decentralised balancing
  of social feedback and noise.
\newblock {\em Swarm Intelligence}, 13(3--4):321--345, 2019.

\bibitem{Mobilia2003}
Mauro Mobilia.
\newblock Does a single zealot affect an infinite group of voters?
\newblock {\em Physical Review Letters}, 91(2):028701, 2003.

\bibitem{Marvel2012}
Seth~A. Marvel, Hyunsuk Hong, Anna Papush, and Steven~H. Strogatz.
\newblock Encouraging moderation: Clues from a simple model of ideological
  conflict.
\newblock {\em Physical Review Letters}, 109(11):118702, 2012.

\bibitem{Gillespie2013}
Daniel~T Gillespie, Andreas Hellander, and Linda~R. Petzold.
\newblock Perspective: Stochastic algorithms for chemical kinetics.
\newblock {\em The Journal of Chemical Physics}, 138(17):170901, 2013.

\bibitem{Reina:SwInt:2015}
Andreagiovanni Reina, Roman Miletitch, Marco Dorigo, and Vito Trianni.
\newblock A quantitative micro-macro link for collective decisions: The
  shortest path discovery/selection example.
\newblock {\em Swarm Intelligence}, 9(2--3):75--102, 2015.

\bibitem{Reina:RAL:2017}
Andreagiovanni Reina, Alex~J. Cope, Eleftherios Nikolaidis, James A.~R.
  Marshall, and Chelsea Sabo.
\newblock {ARK: Augmented Reality for Kilobots}.
\newblock {\em IEEE Robotics and Automation Letters}, 2(3):1755--1761, 2017.

\end{thebibliography}
\end{document}